\documentclass[nonacm]{acmart}

\usepackage{graphicx}

\AtBeginDocument{%
  \providecommand\BibTeX{{%
    \normalfont B\kern-0.5em{\scshape i\kern-0.25em b}\kern-0.8em\TeX}}}

\setcopyright{acmcopyright}

\begin{document}

\title[Development, Deployment, and Evaluation of DyMand]{Development, Deployment, and Evaluation of DyMand — An Open-Source Smartwatch and Smartphone System for Capturing Couples' Dyadic Interactions in Chronic Disease Management in Daily Life}

\author{George Boateng}
\email{gboateng@ethz.ch}
\affiliation{%
  \institution{ETH Z{\"u}rich}
  \city{Zurich}
  \country{Switzerland}
}

\author{Prabhakaran Santhanam}
\email{psanthanam@ethz.ch}
\affiliation{%
  \institution{ETH Z{\"u}rich}
  \city{Zurich}
  \country{Switzerland}
}

\author{Elgar Fleisch}
\email{efleisch@ethz.ch}
\affiliation{%
  \institution{ETH Z{\"u}rich}
  \city{Zurich}
  \country{Switzerland}
}
\affiliation{%
 \institution{University of St. Gallen}
 \city{St. Gallen}
 \country{Switzerland}
}

\author{Janina L{\"u}scher}
\email{janina.luescher@psychologie.uzh.ch}
\affiliation{
    \institution{University of Z{\"u}rich}
    \city{Zurich}
    \country{Switzerland}
}

\author{Theresa Pauly}
\email{theresa.pauly@psychologie.uzh.ch}
\affiliation{
    \institution{University of Z{\"u}rich}
    \city{Zurich}
    \country{Switzerland}
}

\author{Urte Scholz}
\email{urte.scholz@psychologie.uzh.ch}
\affiliation{
    \institution{University of Z{\"u}rich}
    \city{Zurich}
    \country{Switzerland}
}

\author{Tobias Kowatsch}
\email{tkowatsch@ethz.ch}
\affiliation{%
  \institution{ETH Z{\"u}rich}
  \city{Zurich}
  \country{Switzerland}
}

\affiliation{%
 \institution{University of St. Gallen}
 \city{St. Gallen}
 \country{Switzerland}
}

\renewcommand{\shortauthors}{Boateng et al.}

\begin{abstract}
Dyadic interactions of couples are of interest as they provide insight into relationship quality and chronic disease management. Currently, ambulatory assessment of couples’ interactions entails collecting data at random or scheduled times which could miss significant couples’ interaction/conversation moments. In this work, we developed, deployed and evaluated DyMand, a novel open-source smartwatch and smartphone system for collecting self-report and sensor data from couples based on partners’ interaction moments. Our smartwatch-based algorithm uses the Bluetooth signal strength between two smartwatches each worn by one partner, and a voice activity detection machine-learning algorithm to infer that the partners are interacting, and then to trigger data collection. We deployed the DyMand system in a 7-day field study and collected data about social support, emotional well-being, and health behavior from 13 (N=26) Swiss-based heterosexual couples managing diabetes mellitus type 2 of one partner. Our system triggered 99.1\% of the expected number of sensor and self-report data when the app was running, and 77.6\% of algorithm-triggered recordings contained partners’ conversation moments compared to 43.8\% for scheduled triggers. The usability evaluation showed that DyMand was easy to use. DyMand can be used by social, clinical, or health psychology researchers to understand the social dynamics of couples in everyday life, and for developing and delivering behavioral interventions for couples who are managing chronic diseases.
\end{abstract}



\maketitle

\section{Introduction}
Romantic relationships have powerful effects on people’s mental and physical health (see \cite{robles2014} for an overview). For instance, conflicts and negative qualities of one’s intimate relationship are associated prospectively with morbidity and mortality \cite{loving2013}. Romantic or social relationships play an important role in illness management if partners share the responsibility and consider the disease to be their joint problem instead of being only the problem of the afflicted partner  \cite{seidel2012, rintala2013} and it can involve social support and common dyadic coping (CDC) \cite{bertschi2021}. Social support entails providing resources to help a receiver cope in a time of need and can be emotional (e.g., providing comfort or encouragement) or instrumental (e.g., help with practical problems and tasks; \cite{cohen2000, schwarzer2010, luescher2018}). CDC is the "we approach" to dealing with stressors in a couple's relationship \cite{bodenmann2005} which can be assessed objectively by counting first-person plural pronouns \cite{rohrbaugh2008}, by questionnaire or behavioral observation. Social support among couples and CDC in chronic disease management have been shown to have mostly positive effects on emotional well-being \cite{iida2010, bolger2007, rottmann2015, weitkamp2021}, and result in healthier eating habits among diabetes patients \cite{miller2005}. Consequently, it is of interest to better understand couples’ dyadic interactions in-situ, for example, in couples’ management of diabetes in daily life \cite{khan2013, luescher2019} as they could enable the development and delivery of behavioral interventions to, for example, improve physical activity, diet, and medication adherence. 

Ubiquitous devices such as smartphones and smartwatches provide a good opportunity to collect relevant data such as sensor and self-report data from couples in daily life. Smartwatches in particular could be leveraged to collect data on couples’ dyadic interactions and chronic disease management. Several features of smartwatches make them uniquely positioned for this task. Firstly, they are mostly with the wearer since they are worn on the wrist in comparison with a smartphone which could be in various places like the pocket, or bag, and just not in proximity to the user, or devices like Amazon Echo or Google Home which can only be in one place and not always around the owners. Additionally, commercial smartwatches could be used to collect a wide variety of sensor data such as audio and heart rate (for stress detection, emotion recognition), Bluetooth (for proximity detection), accelerometer, and gyroscope (for gestures and physical activity), and ambient light (to detect context). Finally and importantly, smartwatches could be leveraged in novel ways to capture dyadic interactions of partners (e.g., triggering data collection when partners are close and speaking) as we do in this work.

In this work, we describe the development, deployment, and evaluation of DyMand, a novel open-source smartwatch \footnote{\url{https://bitbucket.org/mobilecoach/dymandwatchclient/src/master/}}, and smartphone \footnote{\url{https://bitbucket.org/mobilecoach/dymand-mobilecoach-client/src/master/}} system for ambulatory assessment of couples' \textbf{Dy}adic \textbf{Man}agement of chronic \textbf{d}iseases in daily life.  The DyMand system collects self-report and sensor data based on partners’ interaction moments. In particular, we developed a smartwatch-based algorithm that uses the Bluetooth signal strength between two smartwatches each worn by one partner, and a voice activity detection machine-learning algorithm to infer that the partners are interacting, and then to trigger data collection. We deployed the system in a field study with heterosexual couples in Switzerland that are managing type 2 diabetes (T2DM) of one partner, a common chronic disease affecting 6\% of the Swiss population \cite{idf}. The specific use case in this work was to collect data to understand the association between multimodal sensor data and self-report data of social support, and CDC in the context of diabetes management. This understanding will provide a sound basis for theory- and evidence-based development of dyadic interventions in the context of couples’ dyadic illness management. The DyMand system can be used by, for example,  social psychologists to understand the social dynamics of couples in everyday life and their impact on relationship quality, and also by clinical or health psychologists for developing and delivering behavioral interventions for couples who are managing chronic diseases. This work builds upon a study protocol published in 2019 \cite{luescher2019} and it is an extension of our prior work \cite {boateng2019b, boateng2020d} and it includes a more detailed description of our DyMand system and its real-world deployment and evaluation.

This paper is organized as follows. Next, we discuss related work (Section \ref{sec:related_work}). We then describe the system design (Section \ref{sec:development_design}) and its  implementation (Section \ref{sec:development_implement}). In Section \ref{sec:deployment}, we describe the deployment of the system in a user study and, in Section \ref{sec:evaluation}, we evaluate its technical performance and usability. Finally, we describe limitations and future work in Section \ref{sec:limitations_future} and conclude this work in  Section \ref{sec:conclusion} with a summary.

\section{Related Work}
\label{sec:related_work}
Various smartphone applications have been developed for ambulatory data collection by social and health psychologists. For example, the Electronic Activated Recorder (EAR) has been used in several studies  \cite{mehl2012, robbins2011}, especially for the collection of audio data in various couples’ interactions such as couples managing breast cancer\cite{robbins2014,karan2017}. The EAR triggers data collection at random or scheduled times in the day and collects snippets of ambient sound (e.g., 50 seconds every 9 minutes), which are later transcribed and coded. The EAR does not collect self-report data. On the other hand, a mobile and wearable system was used to collect sensor and self-report data for conflict detection among couples \cite{timmons2017, timmons2017b}. Similar to the EAR, audio data were collected at random or scheduled times in the day (3 minutes of audio every 12 minutes). Another work \cite{reblin2018} used a digital recorder for a whole-day recording of couples managing cancer.

Despite these advances in the ambulatory assessment of couples' interactions, there are still gaps. Firstly, any random or scheduled triggering of data collection does not take advantage of the dyadic nature of couples interactions (e.g., by inferring if partners are actually interacting) and could miss key conversations/interaction moments. The EAR collects only audio and does not leverage potentially more information from other sensor data or self-reports. The all-day recording of \cite{reblin2018} has significant privacy concerns. Consequently, there is currently no ubiquitous system that leverages the dyadic nature of couples’ interaction for the collection of sensor and self-report data that are relevant for social interactions and chronic disease management in everyday life.

\section{Development: System Design}
\label{sec:development_design}
In developing DyMand, experts from the field of computer science, information systems, and health psychology used justificatory knowledge from prior work \cite{miller2005, iida2010, timmons2017, boateng2016, luescher2019} about social support, CDC, health behavior, and emotional well-being to derive a list of design specifications that are important for collecting corresponding data in-situ, in the context of chronic disease management. We describe the specifications.

\subsection{Physical Closeness Monitoring}
The system should track the physical closeness of the partners during waking hours. This information can be used to infer how much time romantic partners spend in various forms of daily interactions (engaging in a shared activity, talking, and arguing), which could be used to predict the couple’s relationship outcomes \cite{hogan2021}. Furthermore, physical closeness tracking can be used in combination with other kinds of data to capture moments when partners are interacting, which would enable the collection of data relevant to chronic disease management such as emotional well-being, social support, and CDC based on partners’ interaction context.

\subsection{Multimodal Sensor Data Collection}
The system should collect relevant 5-minutes worth of multimodal sensor data (in particular audio) once each hour during waking hours (set by couples), ideally when partners are interacting. We restricted the data collection requirement so that we collect only one 5-minute sample per hour (audio data) for privacy reasons to ensure that we do not collect too much audio of the couples’ daily life. Furthermore, the requirement for collecting this data when partners are interacting ensures there is a high likelihood of capturing conversations between partners as compared to collecting data at a random or scheduled time. Sensor data such as audio, heart rate, gestures, physical activity, and step count could be used to manually and automatically infer behavioral information such as social support, CDC, emotional well-being, and health behavior, which are relevant for chronic disease management \cite{luescher2019}. Audio for example can be used to code constructs such as emotions, social support, and CDC. Also, audio together with other data such as heart rate and movements data could be used to automatically detect the emotions of each partner \cite{alhanai2017}. 

\subsection{Self-report Data Collection}
The system should collect self-report data immediately after sensor data collection and at the end of the day. This requirement ensures that various validated self-reported instruments also collect relevant data about social support, CDC, emotional well-being, and health behavior. Collecting this data right after sensor data collection enables matching the responses to the inferences made from the sensor data based on the partners’ interactions. Additionally, collecting this data at the end of the day enables relevant daily summative data such as whether the patient took their medication for the day.

\section{Development: System Implementation}
\label{sec:development_implement}
In this section, we describe our implementation of the system requirements to develop the DyMand system consisting of an overview of the DyMand system, devices used, the MobileCoach platform, smartphone app, and smartwatch app.

\subsection{Overview of the DyMand system}
\begin{figure}
\includegraphics[width=\linewidth]{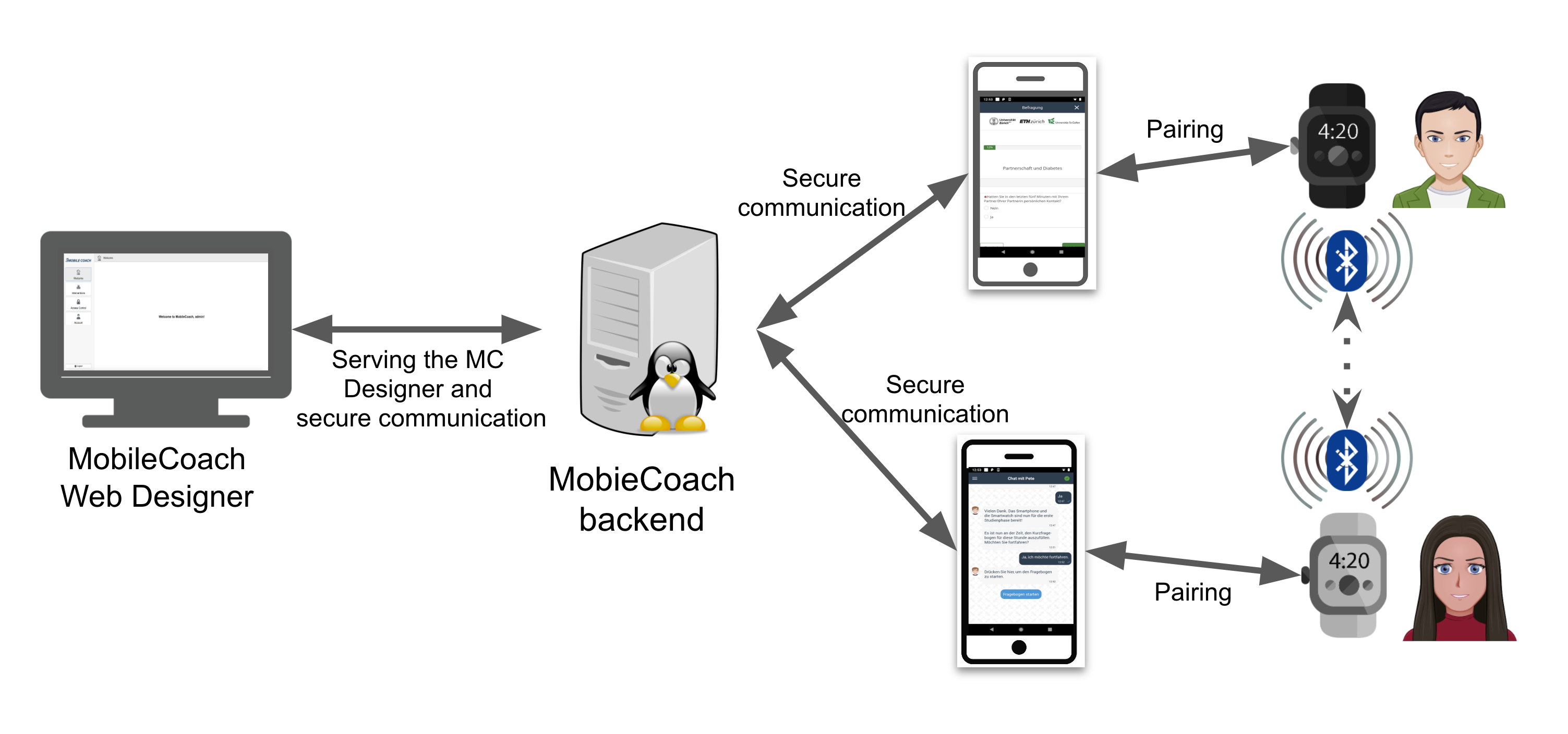}
\caption{Overview of the DyMand system} 
\label{fig:dymand_system}
\end{figure}

The DyMand system (Figure~\ref{fig:dymand_system}) consists of a smartwatch app (section \ref{sec:watch_app}), and a smartphone app (section \ref{sec:phone_app}) built on top of the MobileCoach platform (section \ref{sec:mobilecoach}) \cite{filler2015, kowatsch2017} that consists of a web-based intervention designer and backend. Each partner is given a smartwatch and a smartphone (both paired) running the DyMand apps.

\begin{figure}
\includegraphics[width=\linewidth]{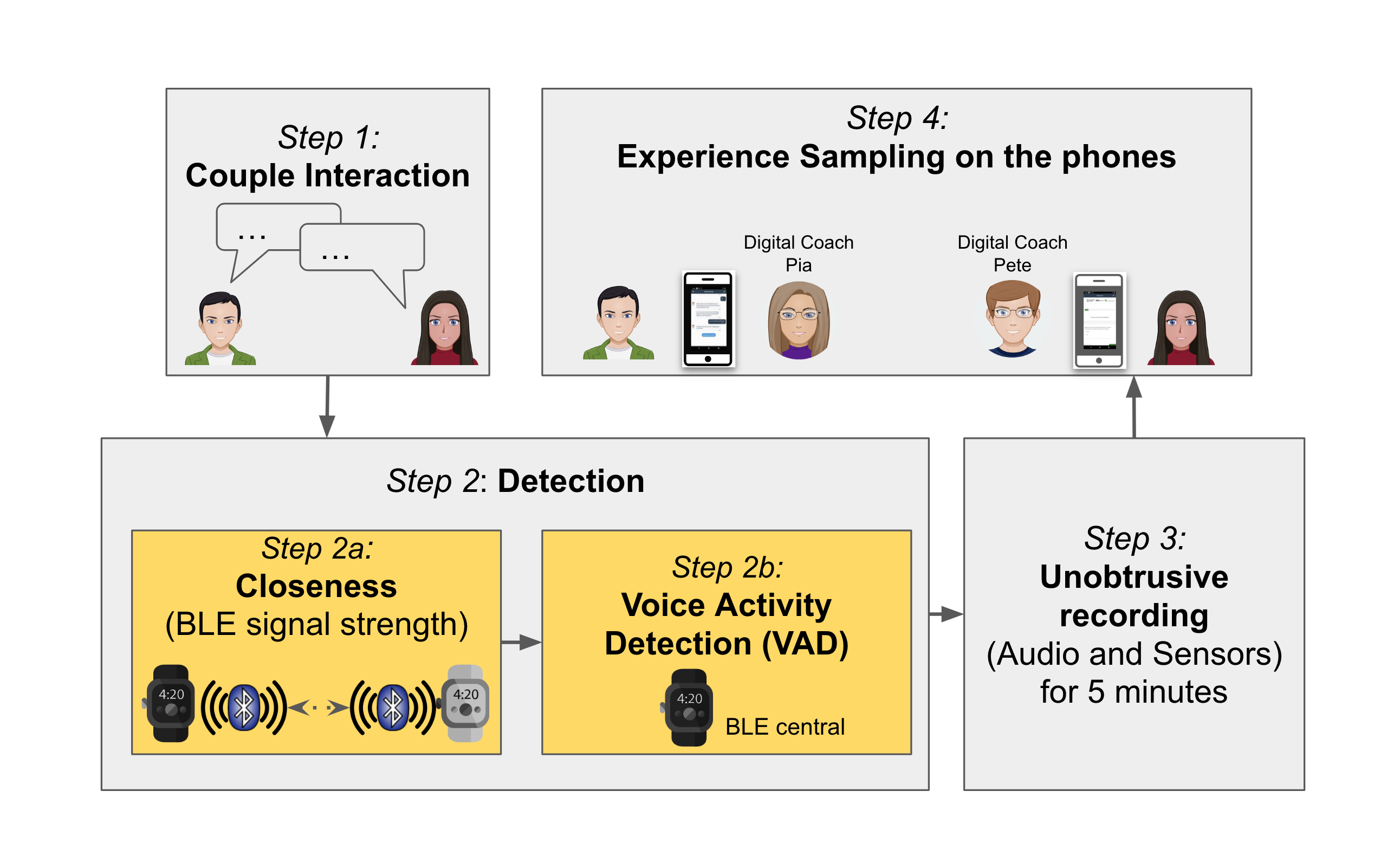}
\caption{Use case of DyMand system} 
\label{fig:dymand_system_use_case}
\end{figure}

The smartwatch app collects 5-mins of sensor data (section \ref{sec:watch-data-collection}) based on the detection of the partners’ interactions via Bluetooth Low Energy (BLE) and Voice Activity Detection (VAD) (Figure~\ref{fig:dymand_system_use_case}). After data collection, it gives a vibration alert on the smartwatch, and then sends an intent to the smartphone app to also give an alert (push notification) and trigger the self-report for each partner to complete separately. The smartphone app then sends a signal to the MobileCoach backend (section \ref{sec:mc-designer}) to trigger the showing of the self-report on the smartphone. The smartphone app is customized with two digital coaches  PIA (interacting with the partner with diabetes) and PETE (interacting with the partner without diabetes) that send various messages to each partner (e.g., “it is time to complete the self-report”). MobileCoach is also used for collecting data from the partners during the setup phase of the devices and also triggers end-of-the-day diary questionnaires. MobileCoach also sends reminders and escalation messages when the expected number of self-reports is not completed.

\subsection{Devices: Polar M600 Smartwatch and Nokia 6.1 Smartphone} 
For implementing the data collection from partners during their conversation moments, we decided to use two smartwatches (one for each partner) as they are more likely to be with wearers often compared to a smartphone, and could better capture physical closeness and interactions. Furthermore, a smartwatch has a better chance of capturing conversations if audio is recorded on it than a smartphone that may not be in the proximity of the partners. Additionally, relevant data of interest such as heart rate and gestures can be captured on a smartwatch and not a smartphone. 

We started working with Apple’s WatchOS platform but to the best of our knowledge at the time of development, we could not find a solution to start an audio recording as a  background process. Hence we moved on to use smartwatches with the Wear OS (based on Android) platform because it provided a lot of flexibility for data collection. We chose a device from the company, Polar as their devices are regularly used in research \cite{polar-research} and we found the Polar M600 smartwatch to be practically priced and the battery seemed to last well with continuous sensing and with a daily charge cycle. Additionally, the heart rate sensor had good performance in a comparison study in which it was shown to be accurate during periods of steady-state activities like cycling, walking, jogging, and running, but less accurate during some exercise intensity changes \cite{horton2017}. 

We decided to use a smartphone to collect self-report data as the screen size of the watch was too small for filling out self-reports, and to serve as an intermediary between the MobileCoach server and the smartwatch app. We used two Nokia 6.1 smartphones running Android 9.0 — one per partner — which we outfitted with SIM cards for Internet access. We found Nokia 6.1 to be practically priced, with a moderate screen size (5.5”) and with good haptics.

\subsection{MobileCoach Platform} 
\label{sec:mobilecoach}
We implemented the DyMand system on top of MobileCoach, an open-source software platform for the design of behavioral interventions and ecological momentary assessments \cite{filler2015, kowatsch2017}. MobileCoach provides a server setup and a mobile client app. The server offers a web-based intervention designer, called MobileCoach Designer (section \ref{sec:mc-designer}). The intervention author can use this interface to design dialogues that can be sent to the mobile app. The author can also put together rules in the designer and connect them to these dialogues so that the messages can be sent when the conditions in these rules are satisfied. The rules could be based on intent from the user (e.g., button press in the mobile app), time range (e.g., between 2 and 3 pm), and other variables which can be created and used in the designer. When the mobile app is opened for the first time, it can show information to the user about the app and let the user pick a conversational agent, i.e. a computer program that imitates the conversation with a human being \cite{kowatsch2021}. By default, the app provides two conversational agents and after one is chosen, the agent can communicate with the user. The agent converses using the dialogues designed in the MobileCoach Designer. With some customizations and a smartwatch app connected with the MobileCoach framework, it fitted the DyMand system’s use case. 

\subsection{MobileCoach Backend}
\label{sec:mc-backend}
\begin{figure}
\includegraphics[width=\linewidth]{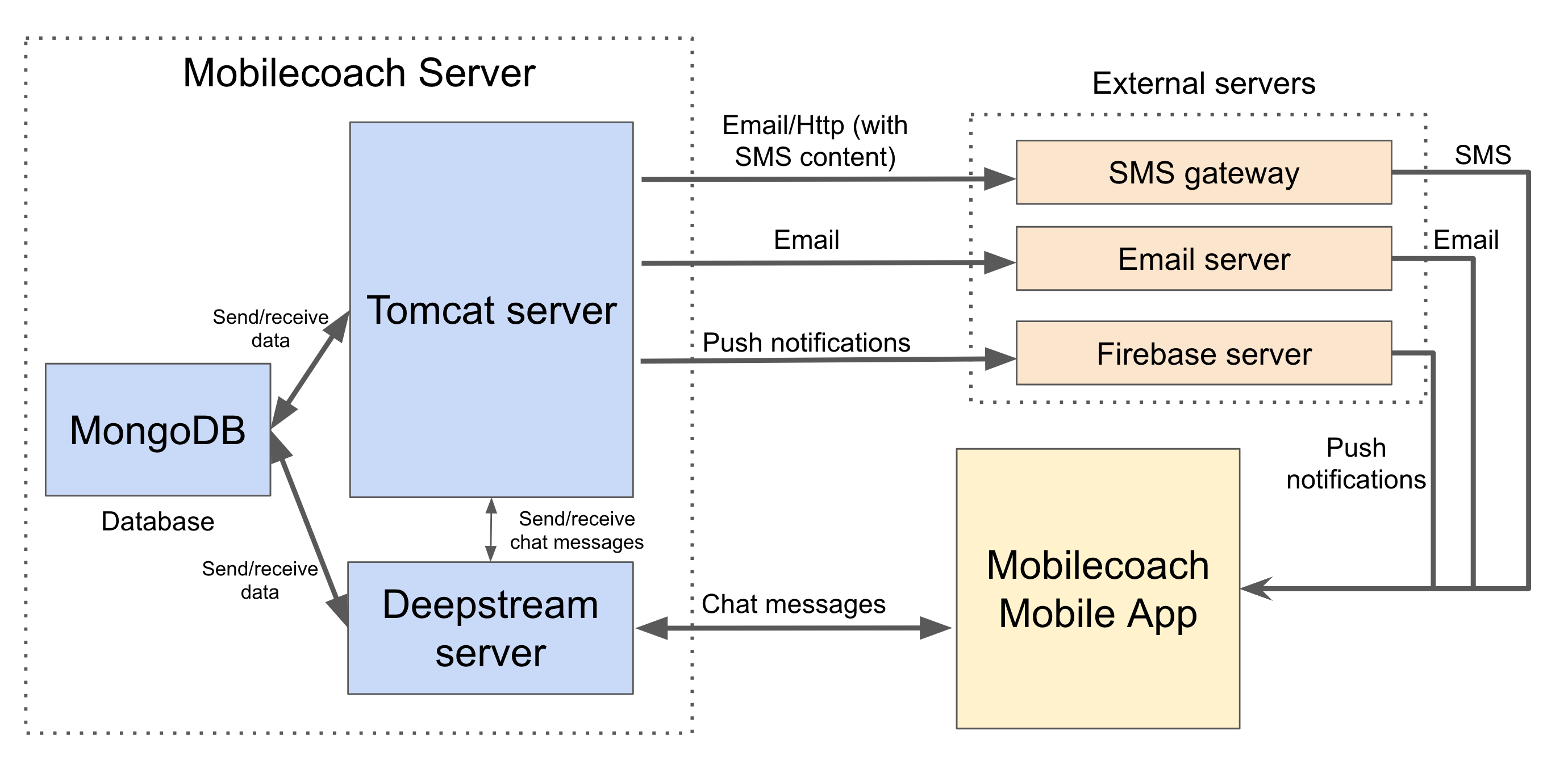}
\caption{DyMand MobileCoach Server and its functions} 
\label{fig:dymand_mc_server}
\end{figure}

The MobileCoach server (Figure~\ref{fig:dymand_mc_server}) of the DyMand system is hosted in an Ubuntu 16.04 virtual machine in the ETH Zurich network. The framework provides a Docker-based setup. The Docker platform \cite{merkel2014docker} performs OS-level virtualization and helps in the portability of software. The server setup of MobileCoach comes with containers running Tomcat \cite{tomcat}, DeepStream \cite{deepstream}, MongoDB \cite{mongodb} and Nginx \cite{nginx}. Tomcat provides a Java web server environment and it serves the MobileCoach Designer (section \ref{sec:mc-designer}) which is built using the Vaadin framework \cite{vaadin}. The intervention engine of MobileCoach processes the rules and schedules the messages. It connects with the user interface (UI) of the MobileCoach Designer and communicates (i.e., sends/receives the messages) with the mobile app using the real-time server DeepStream. DeepStream allows clients and backend services to synchronize data. The database is managed by MongoDB, which is a NoSQL database program. The communication between the server and the mobile app happens through secure SSL \cite{enwiki:1079821520} encrypted connections. We used the certificate authority Let’s Encrypt \cite{letsencrypt} for setting up the free SSL certificates. We performed regular security upgrades and monitored the server. In addition to sending messages via chat, the MobileCoach framework was configured to send SMS or emails. For emails, we used the SMTP server provided by ETH Zurich. For SMS, we used ASPSMS \cite{aspsms} service. Additionally, for the questionnaires, we hosted a LimeSurvey setup in the DyMand server. LimeSurvey \cite{limesurvey} is a free and open-source online survey tool we used to create our questionnaires. We picked LimeSurvey because it provides free community editions and detailed setup instructions, it is easily customizable, and we could host it on our DyMand server so that the data can reside within the ETH Zurich network. 

\subsection{MobileCoach Intervention Designer}
\label{sec:mc-designer}
The MobileCoach intervention designer has user interface (UI) elements (Figure~\ref{fig:dymand_mc_designer}) where dialogs can be created with messages using multiple answer option types. The conversational agent (PIA or PETE) serves these dialogues to the partners. There are three main modes of rule execution in MobileCoach designer as follows: (1) daily execution (once every day at midnight), (2) almost-continuously execution (every 5 minutes) and (3) event-triggered execution (e.g., events can be triggered passively by sensor inputs or manually by user interactions in the mobile application). Each dialogue is connected to either of these rules. When conditions of rules are satisfied, then corresponding dialogues are sent. The conditions are designed based on system and user-defined variables available in the MobileCoach Designer. Each participant has a copy of these variables, which defines the state of the participant. The MobileCoach designer also provides an interface through which study participants and their states (based on the variables) can be monitored. This interface also allows the export of all messages sent and received with timestamps, and the most-recent variable values for each participant for data analyses. For the DyMand system, we designed the following dialogues: 

\begin{figure}
\includegraphics[width=\linewidth]{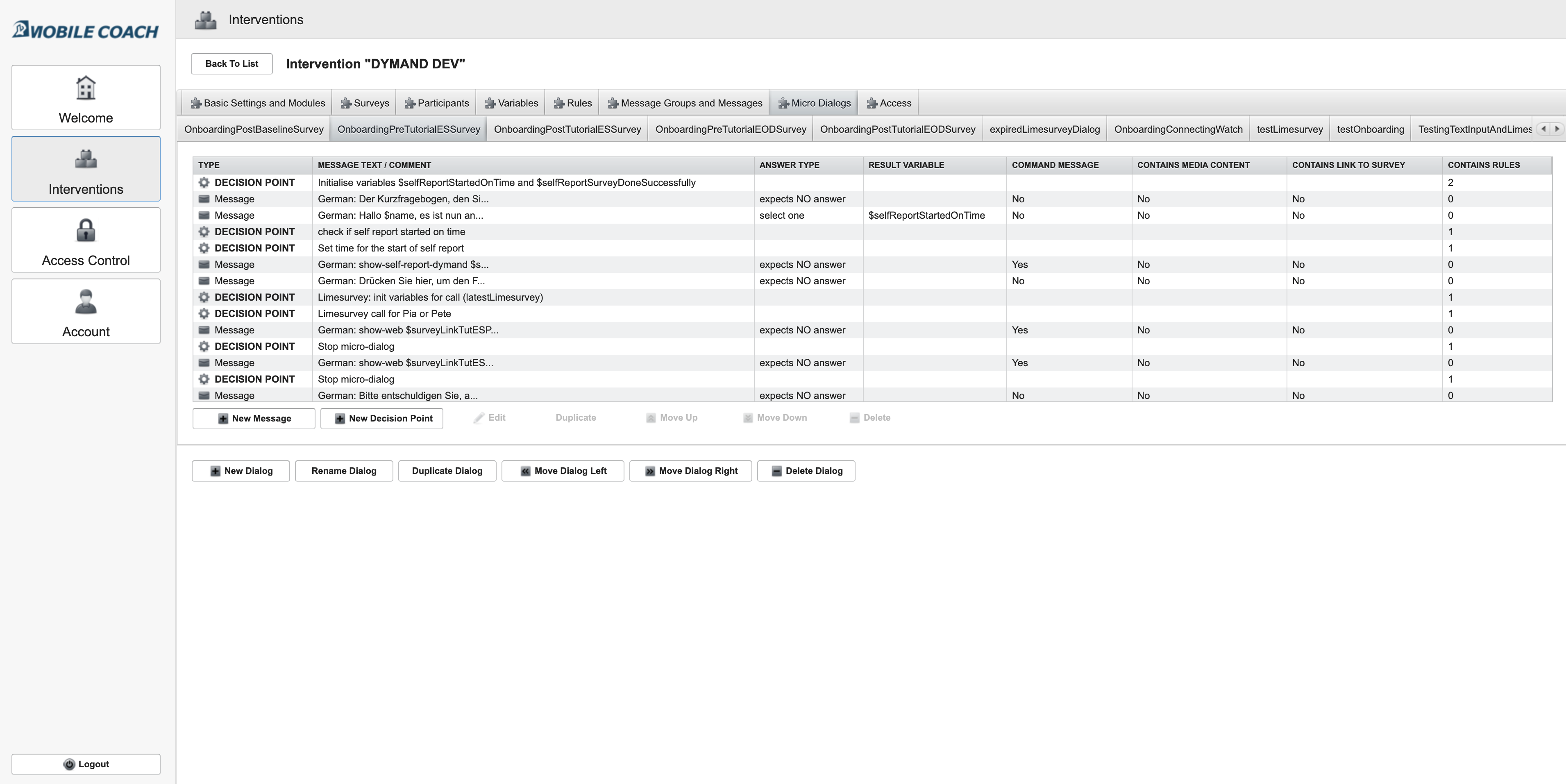}
\caption{MobileCoach Intervention Designer} 
\label{fig:dymand_mc_designer}
\end{figure}

\begin{figure}
\includegraphics[width=\linewidth]{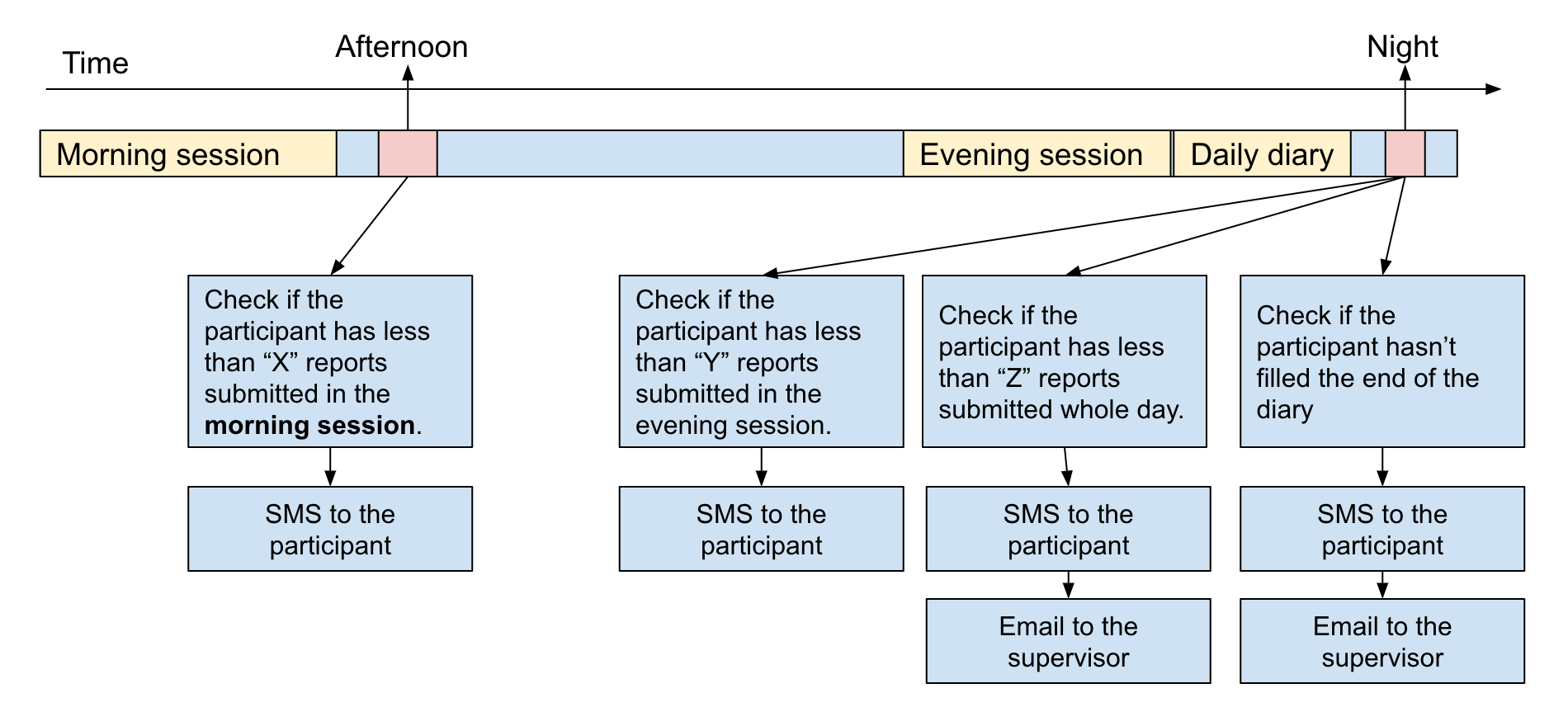}
\caption{DyMand Escalation mechanism: The values X, Y, and Z are calculated every day based on the number of hours the participants specify that they are available for the study in the morning or the evening period and the percentages of self-reports that we set as thresholds for escalation.} 
\label{fig:dymand_escalation_mechanism}
\end{figure}

\begin{enumerate}
\item Onboarding dialogue is sent when the app is opened for the first time during the setup of the study apps. It is used to get the personal information from each partner (e.g., to approach them by their nicknames), to complete the baseline questionnaire before the start of the study, to deliver relevant study information, and help with onboarding the participants for the study.

\item Reminder messages in the form of text messages (SMS) were sent to the personal phones of the couples on Sunday evening and again on Monday morning reminding them to take their study smartphones and wear the study smartwatches directly after getting up to prepare them for the 7-day data collection.

\item Self-report dialogue requests participants to fill a self-report questionnaire after the sensor data collection on the smartwatch. The self-report was a LimeSurvey questionnaire that asked questions about social support, CDC, health behavior, and emotions (e.g., short form of the PANAS self-report \cite{watson1988}). This dialogue used the event-triggered rule execution mode which was initiated when the sensor data collection was successful. In particular, the MobileCoach designer provides various commands to be sent to the mobile app. These commands can have different purposes in the app. For example, a command “show-web <https://abc.de> <Button-name>”, will show a button with the name “Button-name” which when pressed would open the link “https://abc.de” in a new screen within the app. These links in the DyMand system (Figure~\ref{fig:dymand_system_use_case}) were our pre-configured questionnaire from LimeSurvey which additionally provided choices to enter JavaScript code. We added codes in each of our surveys that enabled the app to close the survey screen and continue the dialog with the conversational agent. Using JavaScript code, we added timers to the surveys. They were set to expire after 4 minutes to ensure that partners started filling out the self-report within 4 minutes of getting the alert. Given that the self-report questions were about the previous 5 minutes of the couple's interaction when the sensor data was collected, the time limit helped to ensure the survey data matches the sensor data. 

\item End-of-day diary questionnaire dialogue requests the partners to fill a self-report survey at the end of the day with more comprehensive questions on social support and CDC, healthy eating, medication adherence, and emotional well-being \cite{luescher2019}. Similar to the implementation for the self-report dialogue, they were set to expire after 45 minutes. 

\item Follow-up dialogue was sent when the 7-day field study was finished.  This dialogue asked partners to fill out a follow-up survey about their study experience \cite{luescher2019}. 

\item Escalation messages were designed as both text messages and emails that were sent to the partners and study supervisors when the partners were less adherent to completing self-reports in the study. As shown in Figure~\ref{fig:dymand_escalation_mechanism}, there were two checks done by the system automatically every day during the 7-day study period — one at 2 pm, and another one at the end of the day, after the end-of-day dairy is sent. As shown in Figure~\ref{fig:dymand_escalation_mechanism}, when the partners do not fill at least 60\% percent of the self-reports in the morning or the evening period, a reminder SMS is sent to them. Additionally, if the end-of-day dairy is not completed or if the total number of the self-reports done for the whole day was less than 30\%, then the participant received an SMS and the study supervisor was sent an email additionally to follow up with the couple via phone call. 
\end{enumerate}

\subsection{MobileCoach Smartphone App}
\label{sec:phone_app}
\begin{figure}
\includegraphics[width=\linewidth]{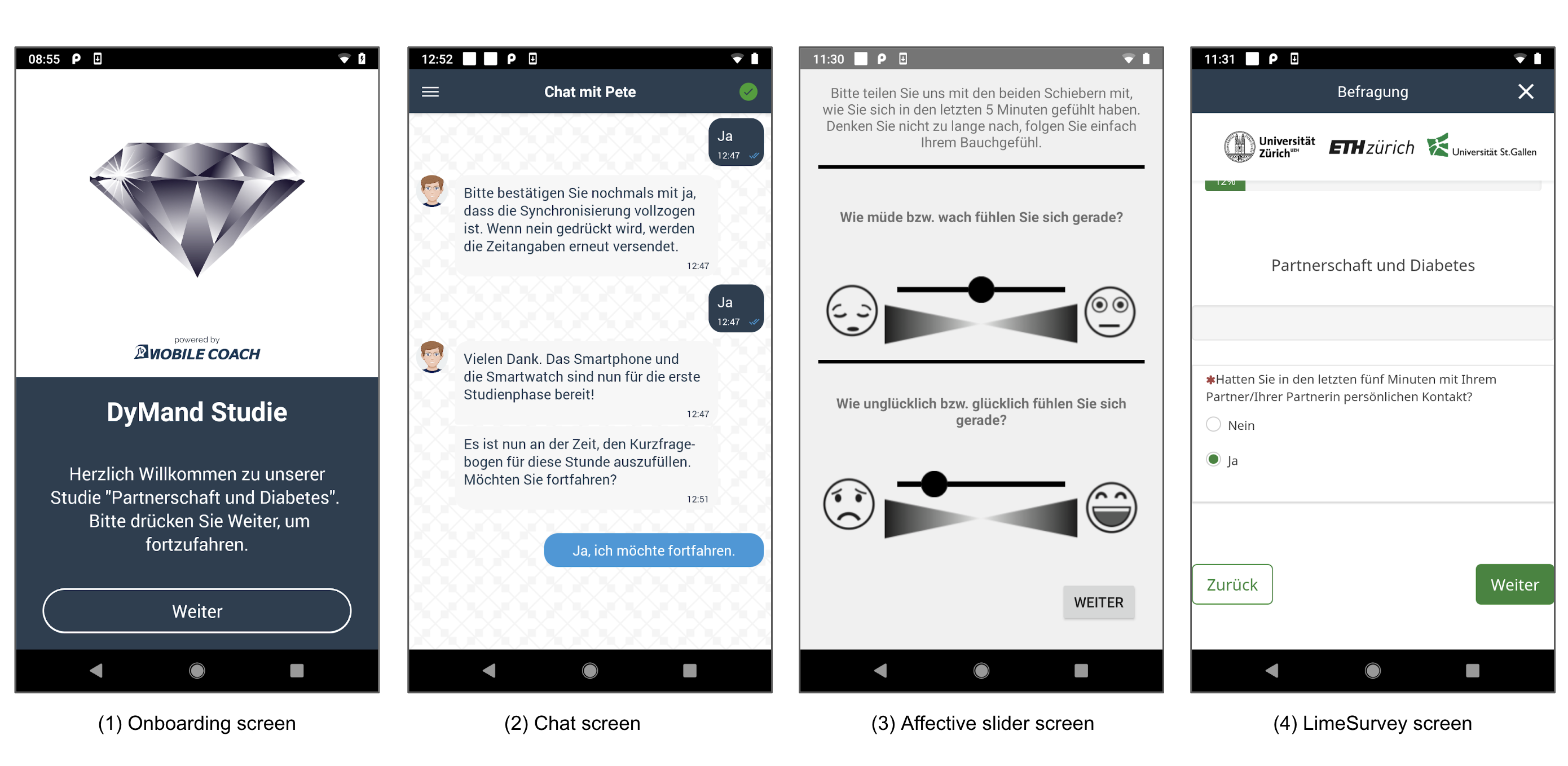}
\caption{Screenshots from the DyMand smartphone app. (1) Onboarding screen: These screens welcome the participants to the study. (2) Chat screen: The digital coach Pia or Pete chats with the participant on this screen. (3) Affective slider screen: The participants can choose their pleasure and arousal levels in the slider and submit them. (4) LimeSurvey screen: The self-reports, the end of the day dairy, baseline, and follow-up surveys are filled by the couples on this screen. } 
\label{fig:dymand_phone_app_self_report}
\end{figure}

MobileCoach provides a skeleton app in React Native, which is a framework for building cross-platform applications for Android and iOS devices. We customized it for the DyMand system and used only the Android app. We customized it to have the two conversational agents PIA (interacting with the partner with diabetes) and PETE (interacting with the partner without diabetes). The smartphone app acts as an intermediary between the smartwatch and the server. It relays the “user intent” messages that indicate that “a recording in the smartwatch has been completed” to the server and also informs the smartwatch as soon as the user has finished completing the self-report triggered by the server. The self-report (Figure~ \ref{fig:dymand_phone_app_self_report}) contains a questionnaire in LimeSurvey \cite{limesurvey} and the Affective Slider, a digital emotion measuring instrument that assesses emotions along the dimensions of valence and arousal \cite{betella2016}. Furthermore, the app collects video, audio, and ambient light for three seconds when each partner is completing the Affective Slider on their smartphone. When a user needs to fill a questionnaire prepared in LimeSurvey, the user sees a button that says “Fragebogen ausfüllen” (Start the survey). When the button is pressed, the link to the survey is opened. The link also includes relevant metadata such as participant code that uniquely identifies the participant, which is needed to link survey answers to the specific partner in LimeSurvey.  

\subsubsection{Continuously Running Smartphone Android App} 
\label{sec:cont-running}
For the DyMand system to function properly, the mobile app is required to run continuously (in the background) on the smartphone. As the smartwatch app was not expected to connect to the Internet on its own, if the connection to the smartphone app was lost, we could not relay the “recording done” message to the server. We implemented a Foreground service \cite{android-fs} in the mobile app which ran continuously and kept the app in a ready state throughout the study. In addition to the Foreground service, we changed the settings on the study phones that we gave to the couples such that the Android OS did not optimize the battery for our DyMand app. Even then, we found that in Nokia 6.1, one of the system apps (com.evenwell.powersaving.g3) still shut down our app after a few hours. We disabled this system app by using the Android Debug Bridge (ADB) and logging into the shell of the device. 

\subsubsection{Smartphone Data Collection} 
When the Affective Slider is shown on the smartphone, a 3-second sensor data recording is made on the smartphone. The recording includes video from the front camera and continuous data from the ambient light sensor. For the video recording from the front camera, we used Android’s  MediaRecorder API \cite{media-recorder-api}. For the ambient light sensor, we used the SensorManager API \cite{sensor-manager-api}. We registered this sensor with the parameter “SENSOR\_DELAY\_FASTEST” to get the sensor data as fast as possible. We stored the data locally on the phone which was retrieved when the couples returned their devices after they finished the study.

\subsubsection{Smartphone and Smartwatch Communication} 
\label{sec:watch-phone-communication}
For the communication between the DyMand apps in the smartphone and the paired smartwatch, we used the Wearable Data Layer API \cite{wearable-data-layer-api}. The following messages were sent between the smartphone and smartwatch applications. 

\begin{enumerate}
\item The weekday and weekend hours during which the couple is available for data collection \cite{wearable-data-layer-api} (chosen by the couple in the smartphone app) are sent from the smartphone app to the smartwatch app during the setup phase (section \ref{sec:setup-watch-phone}). 

\item Text indicating that the smartwatch has finished collecting sensor data for 5 minutes is sent from the smartwatch app to the smartphone app throughout the study. 

\item Text indicating that the self-report on the smartphone has been completed is sent from the smartphone app to the smartwatch app throughout the study.

\item Other messages for acknowledging received messages and logging are sent between the apps on the smartwatch and smartphone. 
\end{enumerate}

\subsection{Smartwatch App}
\label{sec:watch_app}
\begin{figure}
\includegraphics[width=\linewidth]{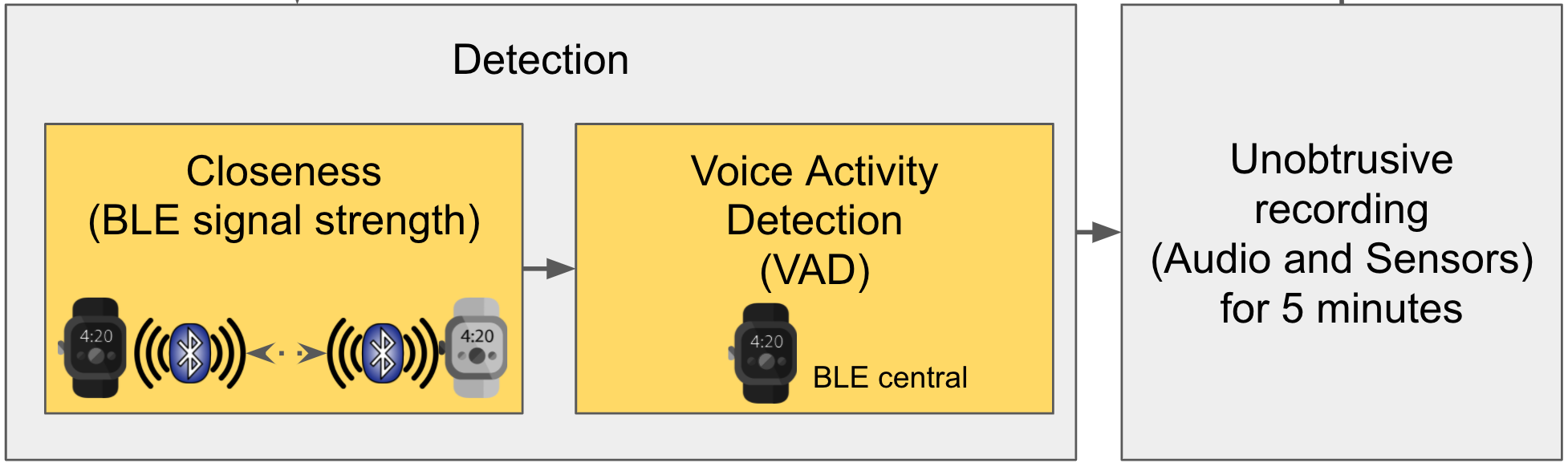}
\caption{Smartwatch App System Overview} 
\label{fig:smartwatch_system}
\end{figure}

Similar to the smartphone app, we implemented a Foreground service \cite{android-fs} in the smartwatch app (Figure~\ref{fig:smartwatch_system}) that collected sensor data. The smartwatch app collected five minutes of the following sensor data once per hour within the morning and evening hours set by the couples: audio, heart rate, accelerometer, gyroscope, and ambient light. We collected a maximum of 5 minutes of data per hour for privacy reasons. Hence, to optimize the quality of data collected within that hour and to ensure that we recorded the most relevant 5 minutes of data (when partners are interacting), rather than triggering data collection at random or scheduled times, the app collected data when 1) the partners were physically close and 2) when there was speech. Our algorithm uses a two-step process. First, the app determines physical closeness using the BLE signal strength between the smartwatches (section \ref{sec:physical_closeness}) and checks if the signal strength is within a certain threshold, which corresponds to a distance estimate (section \ref{sec:ble-service}). Second, the app determines if the partners are speaking by using a voice activity detection (VAD) machine-learning algorithm, which is implemented on the smartwatch (section \ref{sec:vad}). In the case in which this condition of physical closeness and speaking is not met in the hour, the app triggers a backup recording in the last 15 minutes of the hour. 

After the 5-minute recording ends, the watch vibrates and sends a trigger to the smartphone app via the Wearable Data Layer API \cite{wearable-data-layer-api} to bring up the self-report for that partner to complete. If the smartwatch does not receive a message from the smartphone app within 2 mins indicating that the self-report has been started, it gives another vibration alert. If once more, within the next 2 minutes, there is still no response about the start or completion of the self-report, it implies the self-report was not completed. Consequently, the app deletes the audio and attempts to trigger another sensor data collection and self-report for the rest of the hour. Doing this ensured that we collected data with matching sensor and self-report samples. The app also ensured that there were at least 20 minutes between subsequent data collection to reduce the burden of the partners completing the self-reports. 

\subsubsection{Physical Closeness Estimation}
\label{sec:physical_closeness}
\begin{figure}
\includegraphics[width=\linewidth]{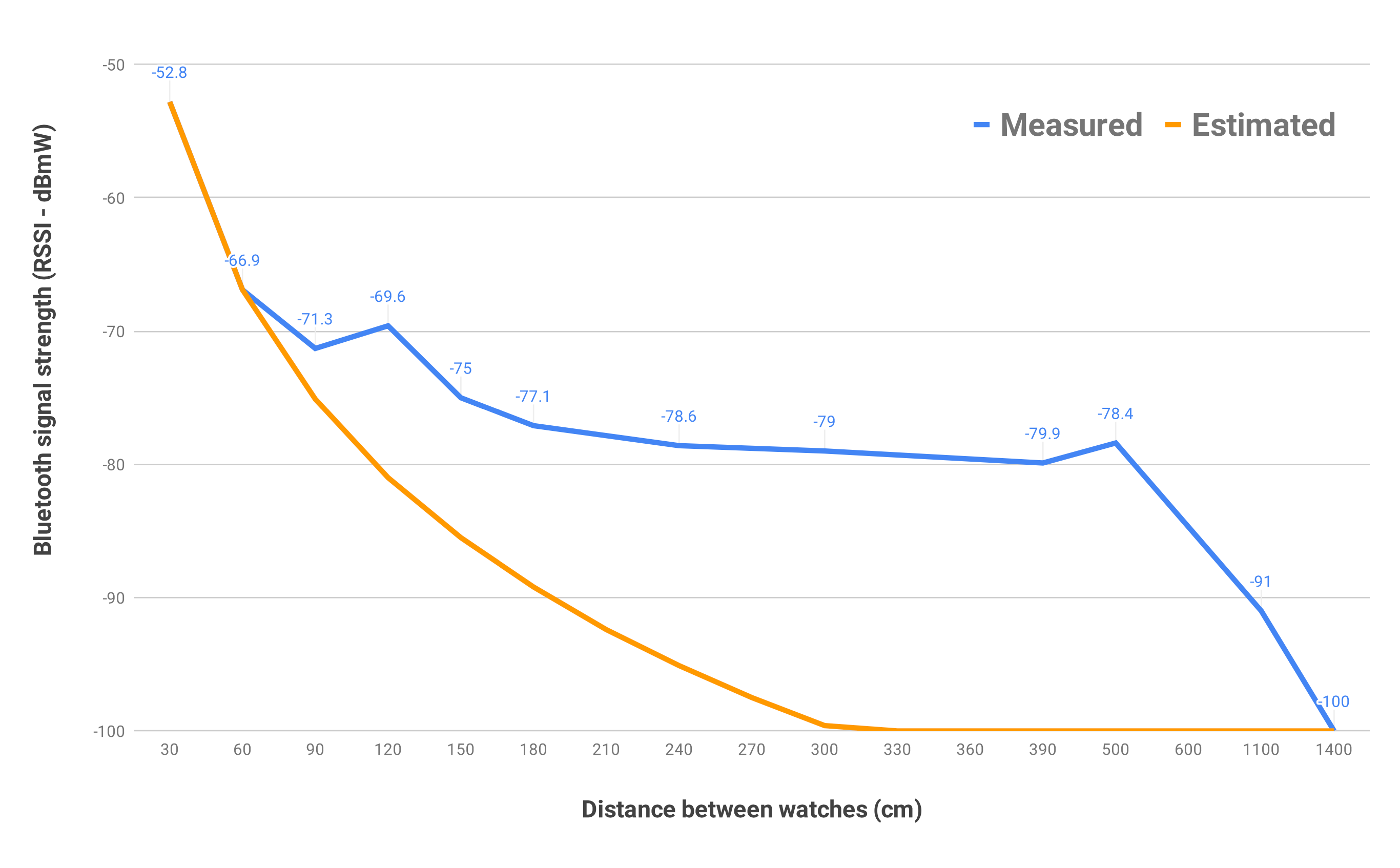}
\caption{Plot of RSSI between two smartwatches versus distance for real-world experiment  and theoretical expectation (that the signal strength increases exponentially when the devices gets closer and closer \cite{jianyong2014})} 
\label{fig:physical_closeness}
\end{figure}

We used the BLE signal strength between the two smartwatches — one acting as the central (does BLE scanning) and the other acting as the peripheral (does BLE advertising) — to estimate the physical closeness of the partners. We conducted an experiment to measure the signal strength based on the distance between the two watches. In a lab setting, we placed two smartwatches at the same level and without any barrier between them. With one BLE peripheral watch fixed at a position, we placed the BLE central watch at a given distance from the former and we measured the signal strength. We varied the distances between the watches and repeated the experiment 10 times. We averaged the values and plotted them as shown in Figure~\ref{fig:physical_closeness} along with the theoretical expectation that the signal strength increases exponentially when the devices get closer and closer \cite{jianyong2014}. Our measurement showed that the signal strength is proportional to the closeness of the smartwatches but not perfect as in any BLE measurement. We chose a threshold of -80dB in the app which covers a range typically less than 5 meters which we assumed should adequately capture the distance between partners when they are interacting. We acknowledge that in the field, the presence of objects such as walls and furniture will affect the signal strength which we did not factor into the experiment. Nonetheless, our goal was not to have a very precise distance versus signal strength mapping but an approximate value to use for closeness between partners.

\subsubsection{Physical Closeness Detection}
\label{sec:ble-service}
For the DyMand smartwatch app, we implemented a BLE service in the smartwatch app for real-time physical closeness detection. At the start of each of the hours assigned for data collection, the peripheral smartwatch continuously advertises its universally unique identifier (UUID). The UUID is created based on the couple ID which is set uniquely for a couple during the onboarding of the study (e.g., P001 for the supporting partner and Z001 for the patient) (section \ref{sec:setup-watch-phone}). The central smartwatch scans with the UUID and when it finds the corresponding peripheral smartwatch, it checks the signal strength. If the signal strength is greater than -80 dB, then it tries to connect to the peripheral smartwatch. After a successful connection, both BLE services acknowledge it by sharing a message through the Bluetooth channel. After this, the central smartwatch does voice activity detection (section \ref{sec:vad}). If the signal strength is less than -80 dB, the BLE scanning waits until the strength breaches the threshold. If there are any problems when the devices connect, the BLE scanning is reset and started from the beginning. Both the central and peripheral BLE services are attached to the Foreground service so that they can run in the background continuously. 

\subsubsection{Voice Activity Detection}
\label{sec:vad}
\begin{figure}
\includegraphics[width=\linewidth]{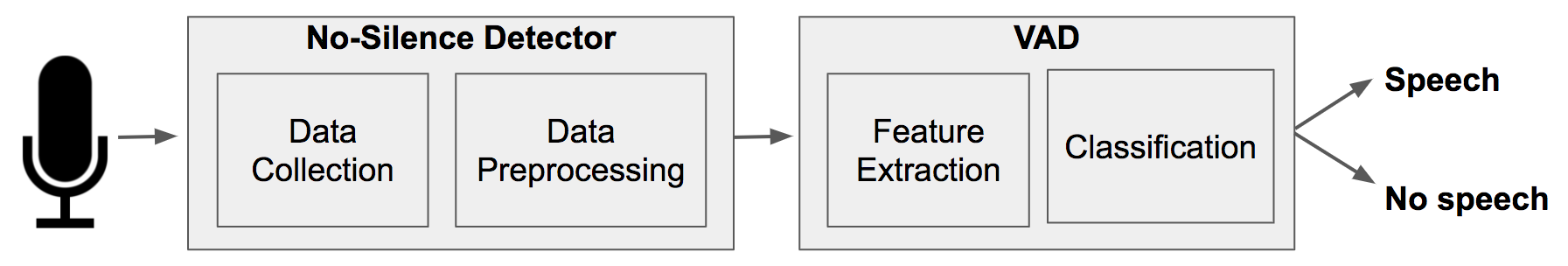}
\caption{System Architecture of VADLite} 
\label{fig:vadlite_system}
\end{figure}

We developed VADLite, a lightweight, open-source voice activity detection (VAD) system that runs in real-time on the smartwatch (see  \cite{boateng2019a} for details of the system) (Figure~\ref{fig:vadlite_system}). An offline and online evaluation of VADLite using real-world data showed better performance than WebRTC's VAD, a popular open-source VAD system. VADLite is a 2-stage system consisting of a no-silence detector and a voice activity detector. 

The no-silence detector computes the root mean square (RMS) of segments of the audio signal and marks them as non-silence if they are above a certain threshold. The voice activity detector consists of a feature extractor, and a machine-learning algorithm, which we trained to classify speech versus non-speech \cite{boateng2019a}. In particular, it extracts mel-frequency cepstral coefficients and classifies speech versus non-speech audio samples using a linear Support Vector Machine (SVM). An SVM is a classifier that constructs a high-dimensional hyperplane to separate data of different classes \cite{hsu2003}. SVM selects a hyperplane that maximizes the distance to the nearest data points on either side of the hyperplane in the case of binary classification. We used a linear SVM because it is memory and computationally efficient when doing predictions. Prior work has used an implementation of linear SVM for real-time prediction on smartwatches for stress detection \cite{boateng2016} and activity detection \cite{boateng2017, boateng2018}. 

\begin{figure}
\includegraphics[width=0.7\linewidth]{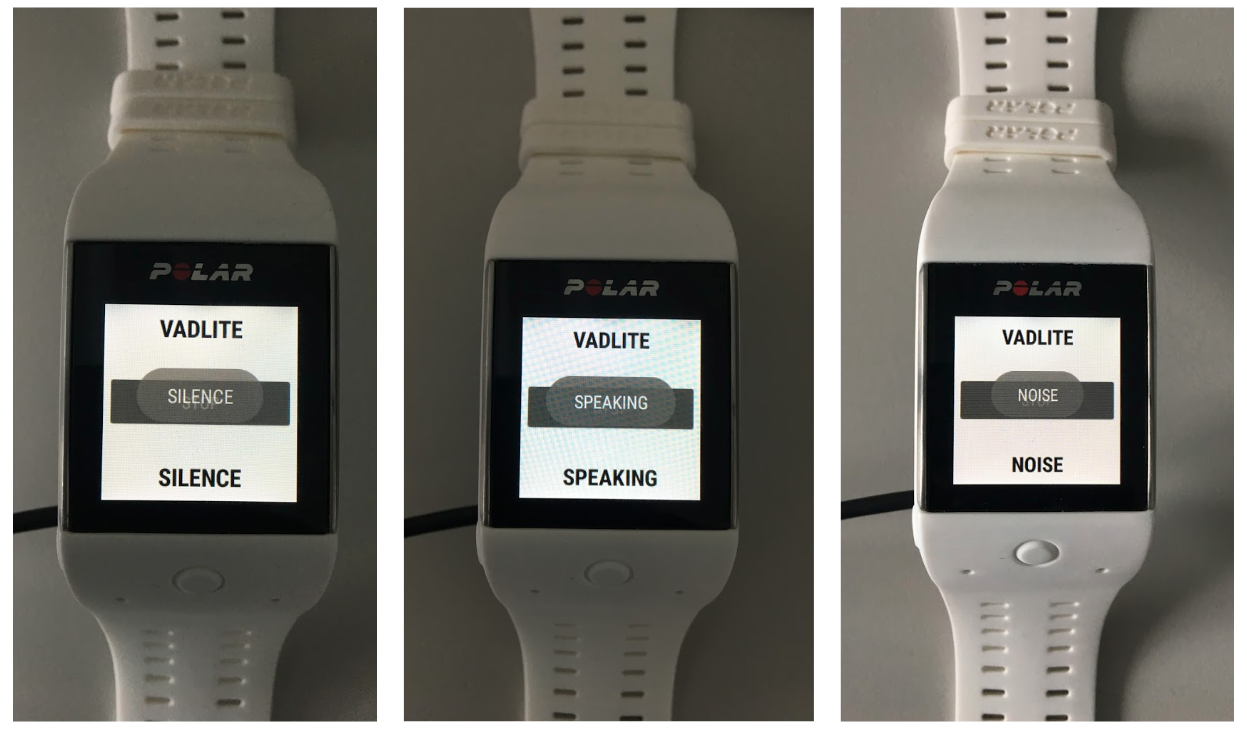}
\caption{Real-time testing of VADLite showing detection states of silence (left), speech (middle) and noise (right)} 
\label{fig:vadlite_realtime}
\end{figure}

To train VADLite, we ran a study that was approved by the ethics commission of ETH Zurich. We collected lab and field audio data (16-PCM mono, 8KHz, 3.5 hours total) using the Polar M600 smartwatch from several people (at least 10 distinct individuals) at varying distances from the smartwatch. We annotated the audio samples as speech or non-speech. We preprocessed the data to remove silence segments by computing the RMS of each one-second time window and checking if the RMS value is below a certain threshold determined empirically. We then extracted frequency-based features  — 13 MFCC features which have been widely used for VAD \cite{forlivesi2018} and used 12 of them (excluding the 1st coefficient, which is the DC component) — over a time window of 25 ms non-overlapping time window. We normalized the features and used them to train a linear SVM to classify speech or non-speech. We then implemented the system to run in real-time on the smartwatch (Figure~\ref{fig:vadlite_realtime}). The real-time implementation continuously collects audio and processes them in 1-second segments for no-silence detection and then voice activity detection.

We performed offline and online evaluations using the following metrics for evaluation: accuracy, speech hit rate (SHR), and false alarm rate (FAR). The SHR is the ratio of correctly detected speech frames to the total number of speech frames. By contrast, FAR is one minus the noise hit rate, where noise hit rate is the ratio of correctly detected noise frames to the total number of noise frames. For offline evaluation, we split the data into train and test using about 70\%-30\% stratified split and performed 10-fold stratified cross-validation with hyperparameter tuning on the train data and evaluated on the test set. The model achieved 82.6\% accuracy, 80.2\% SHR, and 14.9\% FAR. For online evaluation, we played 15-minute audio collected from a naturalistic context through a loudspeaker as the VADLite app performed real-time classification of the audio just like was done by Feng et al \cite{feng2018}. VADLite had SHR and FAR of 91.6\% and 5.5\% respectively. These results were better than WebRTC’s VAD: offline accuracy, SHR and FAR of 71.4\%, 79.5\%, and 37.5\% respectively,  and online  SHR and FAR of 73\% and 18\% respectively. Furthermore, VADLite had an average processing time of 2 ms for each 25 ms frame and 76 ms for the total one-second duration.  As a result, throughput was met since the frame processing time was less than the 25 ms segment duration. Likewise, the processing time for the whole duration was less than one second.

For the DyMand smartwatch app, the VAD component is triggered after the physical closeness detection. It records and samples audio at 8KHz and processes them in 5-second chunks by first performing no-silence detection and then speech detection. If it classifies the segment as speech, the VAD audio recording is stopped, and then sensor data collection is started on the central device (since it runs the VAD component) and a signal is sent to the peripheral device to immediately start sensor data collection.

\subsubsection{Smartwatch Data Collection}
\label{sec:watch-data-collection}
The data collection component of the smartwatch app collected audio, heart rate, accelerometer, and gyroscope for 5 minutes. For the audio data, we used the MediaRecorder API \cite{media-recorder-api} available in Android. We collected 16-PCM mono at 44.1 kHz. We set the output format of the audio file as ".wav” which is a lossless file format. All the other sensor data were collected using the SensorManager API \cite{sensor-manager-api}. We registered these sensors with the parameter “SENSOR\_DELAY\_FASTEST” to get the sensor data as fast as possible. The data collected was stored locally on the smartwatches and retrieved after the devices were returned. 

\subsubsection{Exception Handling}
We added try-catch statements in various parts of the code where they could be exceptions (e.g., writing text or data to a file, BLE scanning, etc.). Additionally, similar to \cite{klugman2018}, we included the “DefaultUncaughtExceptionHandler” to catch all uncaught exceptions. Our implementation of the Exception class spawns a thread that counts the number of exceptions in that hour and saves it (for later logging), saves the exception message (for logging), schedules a restart of the app with an AlarmManager for the next second, stops the current Foreground service, and shuts down the app. 

\subsubsection{Logging}
\label{sec:logging}
To ensure that we understood how well the system was performing, we included various logs in the smartwatch app that were saved in files on the smartwatch. In particular, we had 1) configuration logs at the time of setup, 2) hourly logs between the setup time of the devices and the start of the study, 3) hourly logs during the 7-day study period, 4) continuous log of the BLE signal strength between the two smartwatches, 5) hourly app function logs and 6) error logs as and when they happened. 

\begin{enumerate}
\item The configuration log contained the dates and hours of data collection set by the couples which we used for post-study analysis. 

\item The before-study logs contained the log’s timestamp, battery level, the number of days, hours, minutes, and seconds until the start of the study, and the number of exceptions that happened in the previous hour. We could use this data to infer various things such as whether the battery died before the study started and the partners forgot to charge it, or whether an error shut down the app. 

\item The during-study hourly logs contained several important fields. These included the following from the previous hour: timestamp of the log, battery level, date and number of times the BLE started advertising (peripheral device) or scanning (for the central device), date and number of times the device met the closeness condition, date and number of times of no-silence detection, date, and number of times of voice activity detection, date and number of times the two watches connected, date and number of times sensor data was collected, date and number of times self-report was triggered (1st and 2nd alert), started, and completed, whether the recording was a backup recording, if the audio was discarded (because self-report was not completed), number and dates of errors, number of times and dates the app restarted, whether internet was available on the smartphone, and amount of space remaining on the smartwatch. These data allowed us to assess the performance of the system (see Section \ref{sec:evaluation}). 

\item The hourly app’s function logs contained various “print” statements we had in various functions in our code as well as the Wear OS system logs which our app logged and saved in a file. We did this to debug and better understand which code block may have caused any errors. 

\item The BLE log saved the signal strength between the devices throughout the continuous scanning during the hours of data collection. This data can be used to infer how much time partners spent together. 

\item The error logs were the exceptions (described in the previous section) that we logged. We used those to debug the app.
\end{enumerate}

\subsubsection{DyMand App Checker}
In early deployments, we realized that the smartwatch app could sometimes be shut down by the Wear OS system after an error, and our implementation for the app to restart itself was not always reliable. Consequently, we developed a smartwatch app — DyMand App Checker — to continuously check each hour if the DyMand smartwatch app is running and then start it if it is not running.

\subsection{Setup Components of Smartwatch and Smartphone Apps}
\label{sec:setup-watch-phone}
We implemented setup components of the smartphone and smartwatch apps for the partners to set up their devices, to ensure all aspects of the apps were working correctly before they take the devices away, and also for them to experience the process of sensor and self-report data collection. The smartphone app collects personal data and also the hours that the partners indicate for data collection. This information is transferred to the smartwatch app during the setup of the smartwatch after the “send” button is pressed on the smartphone. If the watch receives it, it shows a screen saying that the configuration process is successful. Otherwise, it needs to be sent from the smartphone again through the same button. The communication between a paired smartwatch and smartphone is explained in detail in Section \ref{sec:watch-phone-communication}. The smartwatch app collects information such as the couple ID (e.g., 001, 002) for logging purposes and creates a unique UUID (section \ref{sec:ble-service}) that can be used by the BLE services for advertising and scanning, the color of their smartwatch for the app to know if the watch belongs to the patient or supporting partner (by default, our smartwatch app chooses white to be the BLE peripheral and the black to be a BLE central device), receives the data collection hours from the smartphone, and collects a voice sample for 1 minute from the partner by asking them to read some text on a paper. Next, the BLE central smartwatch starts scanning and the BLE peripheral smartwatch starts advertising. After successful connection, a message is shared via this Bluetooth connection between the smartwatches, and the recording of the sensor data starts in the watches, completes after 5-minutes, triggers self-report on the smartphone, and then triggers the end-of-day dairy. 

\section{Deployment: Field Study}
\label{sec:deployment}
\begin{figure}
\includegraphics[width=\linewidth]{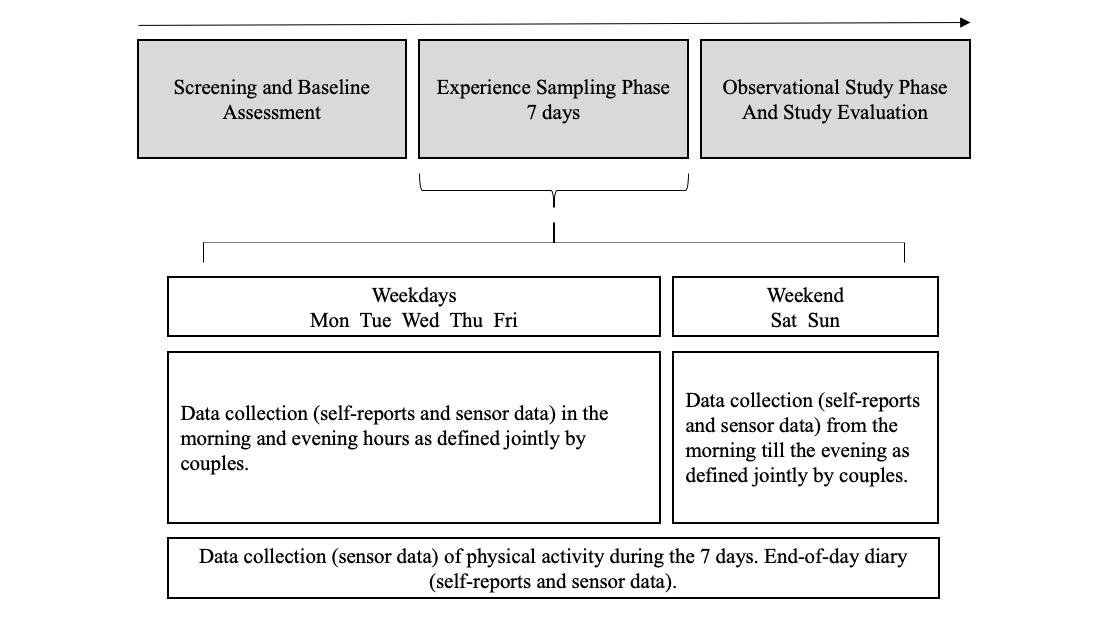}
\caption{Overview of DyMand study \cite{luescher2019}} 
\label{fig:dymand_study}
\end{figure}

After various internal pilot tests of the DyMand system, we deployed it in a field study with couples. We ran the DyMand study between 2019 and 2021 with heterosexual romantic couples from the German-speaking part of Switzerland in which one partner had T2DM (Figure~\ref{fig:dymand_study}) \cite{luescher2019}. In total, we collected data from 13 couples aged 47 to 81 years, with a mean age of 68 (SD = 9) resulting in a total of 1,019 5-minute samples of sensor data (85 hours) and 598 corresponding completed self-report data. 

The study was advertised in various places including hospitals, magazines, local newspapers, and the diabetes association in Switzerland. Interested couples completed a web-based questionnaire to screen them for the inclusion and exclusion criteria, and collect socio-demographic information. Those who met the eligibility criteria were able to pick a date for a baseline assessment at the Applied Social and Health Psychology laboratory of the University of Zurich. During this session, both partners received comprehensive information about the study, signed the informed consent form, and completed a web-based questionnaire that captured constructs of interest at baseline that were not assessed daily. 

They also received instructions on the study and then trained research assistants helped them to set up their devices and pair the corresponding smartphone and smartwatch (see Section \ref{sec:setup-watch-phone} for details of the setup process). Each partner was given a smartwatch and smartphone running the DyMand apps and they were instructed to have all devices with them every day for 7 days from getting up until going to bed. To prevent mistakes from one partner accidentally using the other partner’s watch and phone, one set of phones and watches had black covers and the other set had white covers. The patient was given the white set and the supporting partner was given the black set. The partners picked the hours during which we could collect data from them. They could choose any period from 4 am to 11 am for the morning hours and from 4 pm to 11 pm for the evening hours. During the weekend, only the early morning hours and late evening hours were set (e.g., from 6 am to 10 pm). With this procedure, privacy aspects were addressed by primarily focusing on situations, in which the couples spent time together and thus reducing the number of audio recordings during the day of weekdays when chances are higher that subjects are working or moving around in public places.

We collected data from their daily life for 7 consecutive days starting the next Monday after their visit until the following Sunday night. The DyMand system collected sensor and self-report data as described in previous sections. During the study, we had a process for monitoring the study to ensure that the system was working correctly and to enable us to intervene if needed. We had a spreadsheet with cells corresponding to the hours during which we should have received data. Research assistants checked the DyMand MobileCoach server daily to see if the self-reports were triggered and completed, with each cell receiving one of the following values: 0 — no self-report trigger received,  1 — self-report trigger received, but no survey completed 2 — trigger received and survey completed. Additionally, we had a sheet that we used to track the details of issues and complaints that the couples communicated to the research team either via email or calls. We collected the following relevant information to understand and solve any problems that came up: subject id, device id, does the issue involve the watch, does the issue involve the phone, detailed description of the issues and pictures if available, date and time of issue, where they were when the issue happened, was the watch switched on, was the phone switched on, were the two watches physically close together, were the phone and watch physically close together, was there a connection to the digital coach indicated by the top-right icon on the smartphone app being green, were the phone and watch connected, does the phone have internet,  and are both apps on the watch running.

There are significant ethical and privacy concerns of such a system and study since we collect audio which is sensitive data, and more so in the context of couples’ interactions with the likelihood of speech about private topics. We took several measures as follows. First, our study received ethical clearance from the cantonal ethics committee of the Canton of Zurich, Switzerland (Req-2017\_00430). Second, we ensured that we collected a maximum of 5 minutes of audio per hour in order not to record a significant percentage of the couples’ everyday life. Consequently, even if the system triggers multiple recordings in the hour, the app always deletes all but the last one before the end of the hour. Third, to protect the privacy of subjects not taking part in the study, we asked subjects to wear a tag that we give them to indicate to others around that recording may be happening and that they may be recorded. Finally, after subjects returned their devices, we gave them the option to listen to and request the deletion of any audio samples without any explanation before the study team could listen to the audio files. Similar measures have been used in previous studies \cite{mehl2012, robbins2014} and have proven adequate to safeguard the privacy of study subjects and others not taking part in the study.

\section{Evaluation}
\label{sec:evaluation}
We evaluated the DyMand system regarding its technical performance and usability. For technical performance, the DyMand apps on the smartwatch and smartphone performed hourly logs (section \ref{sec:logging}) of relevant system performance metrics such as whether data collection started and completed, and whether self-report data was triggered, started, and completed. Furthermore, we annotated the audio data with relevant information such as whether it contained speech, and whether there was a conversation between partners for the algorithm triggered recordings and the backup recordings.

\subsection{Self-report and Sensor Data Collection}
We investigated the percentage of the total expected number of sensor data and corresponding self-report data that was collected. Given that couples indicated the hours during which we could collect data, we estimated the total expected number for each couple and then summed them for all couples. Furthermore, given that the DyMand smartwatch app may not always be running due to circumstances such as the device being off because the couples did not charge them or the app crashing because of various errors such as in the BLE stack, a better metric would be the percentage of the expected number of sensor data and self-report data that was collected for hours when the app was running. For example, our logs showed that one couple did not turn on the smartwatch for the duration of the study. Hence, the app did not run and could not collect any data. Using the app’s hourly logs, we estimated the number of hours that the app was running. We then used the following log status data: Was the sensor data collected? Was the self-report triggered? Was the self-report started? and Was the self-report completed. We computed sums of these log events for each partner, and then for all couples (Table \ref{tab:performance_sensor_self_report}). Additionally, we computed relevant percentages as shown in Table \ref{tab:performance_sensor_self_report_perc}. 

Our DyMand system collected 73.2\% of the total expected sensor data and triggered 73.2\% of the total expected self-reports. Considering only the case where the app was running, these percentages become 99.1\% which shows that the system adequately triggered data collection as expected. Additionally, partners started 60.6\% and completed 58.7\% of the triggered self-reports. This percentage is not very high yet understandable for the context of this study. Partners had a maximum of 4 minutes to start the self-report after the trigger. The self-report was dismissed if it was started after this time. There are several reasons partners may not have started the self-report such as they did see or hear the alert because they were not wearing the watch, or they did not have the phone close by them. For those who saw the alert and attempted to start the self-report, some partners complained that there was a delay in the self-report loading on the smartphone due to internet connection issues (see Section \ref{sec:challenges}) and may have affected the start of the self-report in time. 

\begin{table}[]
\centering
\caption{Expected and actual number of sensor and self-report data that were collected}
\label{tab:performance_sensor_self_report}
\begin{tabular}{|l|l|}
\hline
\textbf{Data}  & \textbf{\# of Samples} \\ \hline
Total expected \# of sensor and self-report    & 1392   \\ \hline
Expected \# of samples with the app running  & 1028  \\ \hline
\# of sensor data collected                    & 1019                   \\ \hline
\# of self-reports triggered                   & 1019                   \\ \hline
\# of self-reports started                     &  618                      \\ \hline
\# of self-reports completed                   & 598                    \\ \hline
\end{tabular}%
\end{table}

\begin{table}[]
\centering
\caption{Percentages of the expected number of sensor and self-report data that were collected}
\label{tab:performance_sensor_self_report_perc}
\begin{tabular}{|l|l|}
\hline
\textbf{Data} & \textbf{Percentage (\%)} \\ \hline
\% of total expected sensor data that was collected & 73.2 \\ \hline
\% of expected sensor data with the app running that was collected & 99.1 \\ \hline
\% of total expected self-report triggers that happened & 73.2 \\ \hline
\% of total expected self-report triggers with the app running that happened & 99.1 \\ \hline
\% of triggered self-report that were started & 60.6 \\ \hline
\% of triggered self-report that were completed & 58.7 \\ \hline
\end{tabular}%
\end{table}

\subsection{Capturing Partners’ Conversation Moments}
\label{sec:conv_moments}
Given the key novelty of our system is its touted capability of capturing partners’ conversation moments using physical closeness and voice activity detection, we investigated how well the DyMand system captured speech and conversation between partners. We define a conversation as the presence of speech from both male and female partners in that audio. We had 3 trained research assistants annotate all valid audios (i.e., were not corrupted and hence playable, N=1014) by providing ‘yes’ or ‘no’ to relevant information such as: does it contain speech, did the male partner speak, did the female partner speak, and was there a conversation between partners. We automatically extracted the information about whether each audio was a triggered or a backup recording by looking at the timestamp. Audios collected anytime between the 44th minute and the end of the hour  (e.g., between 6:44 am and 7 am) are backup recordings. All audios collected before then were triggered by our algorithm. We show the sum of audio status information for all partners in Table~\ref{tab:audio_data_stats} and calculate relevant percentages in  Table~\ref{tab:audio_data_stats_perc}. 

\begin{table}[]
\centering
\caption{Number of audio samples with speech and conversation between partners broken down by triggered and backup recordings }
\label{tab:audio_data_stats}
\begin{tabular}{|l|l|}
\hline
\textbf{Data Field} & \textbf{\# of Samples} \\ \hline
Total audios & 1014 \\ \hline
Audios with speech & 791 \\ \hline
\# of triggered recordings & 277 \\ \hline
Triggered recordings with speech & 256 \\ \hline
\# of backup recordings & 737 \\ \hline
Backup recordings with speech & 535 \\ \hline
Conversation between partners for all audios & 538 \\ \hline
Conversation between partners for triggered recording & 215 \\ \hline
Conversation between partners for backup recording & 323 \\ \hline
\end{tabular}%
\end{table}

\begin{table}[]
\centering
\caption{Percentage of audio samples with speech and conversation between partners broken down by triggered and backup recordings }
\label{tab:audio_data_stats_perc}
\begin{tabular}{|l|l|}
\hline
\textbf{Data} & \textbf{Percentage (\%)} \\ \hline
\% of total audios with speech & 78 \\ \hline
\% of triggered recordings with speech & 92.4 \\ \hline
\% of backup recording with speech & 72.6 \\ \hline
\% of total audios with conversation between partners & 53.1 \\ \hline
\% of triggered recordings with conversation between partners & 77.6 \\ \hline
\% of triggered recordings where one partner spoke & 88.1 \\ \hline
\% of backup recordings with conversation between partners & 43.8 \\ \hline
\end{tabular}%
\end{table}

For triggered audios, 92.4\% contained speech, which shows VADLite has good performance in capturing speech in the real world considering prior work has shown that open-source VAD tools perform poorly on real-world speech data collected with smartwatches \cite{boateng2019a, liaqat2018}. Furthermore, 77.6\% of triggered recordings had a conversation between partners in comparison to 43.8\% of backup recordings which had a conversation between partners. This result shows that our novel approach of physical closeness plus speech detection was better at capturing couples conversation/interaction moments (77.6\%) in comparison to using random or scheduled times per hour in our study (43.8\%). It is difficult to have a direct comparison to other works. Nonetheless, in the following work \cite{sels2019b}, random triggering (10 times a day for one week) of self-report data collection among couples resulted in data for which partners were together (a proxy for interaction) in 39\% of the triggers. Also, another work that used the EAR triggered data collection among couples at a scheduled time sequence (every 9 minutes during waking hours) resulting in data for which patients and spouses talked (a proxy for partners’ speech), on average, 47.9\% and 45.0\% of their waking hours \cite{robbins2014}. These percentages are lower than the 77.6\% from the DyMand system.

There are some likely reasons for the absence of conversation between partners in triggered recordings — the outstanding 22.4\%. Firstly, our definition and evaluation metric of conversation between partners is strict — both male and female partners spoke in that 5-minute audio. In actuality, the partners may have been having a conversation but one partner may not have spoken in the specific 5-minute period in which we collected data, resulting in a “no” for conversation between partners. When we relax that definition to be either partner spoke, the percentage increases to 88.1\%. Secondly, our physical closeness detection approach assumes that the partners are always wearing the smartwatch which was not always the case. For example, if the two smartwatches are left together on a table with the radio or TV in the background, a recording was triggered but it did not contain a conversation between partners. One way to address this issue is to include a pre-step to estimate if the device is being worn using accelerometer or heart rate data as an example. Thirdly, another edge case that happened was partners sitting together and watching TV but not having a conversation. In this case, the recording was triggered because of the physical closeness and speech from the TV but the audio did not contain speech from the partners. One way to address this issue is to use an extra step for speaker identification to check if the speech is from either partner. Doing this will entail collecting a voice sample from both partners at setup, and then automatically creating and storing an acoustic fingerprint of each partner on the watch, and checking if there is a match after the voice activity detection.

\subsection{Usability Results}
\begin{figure}
\includegraphics[width=0.7\linewidth]{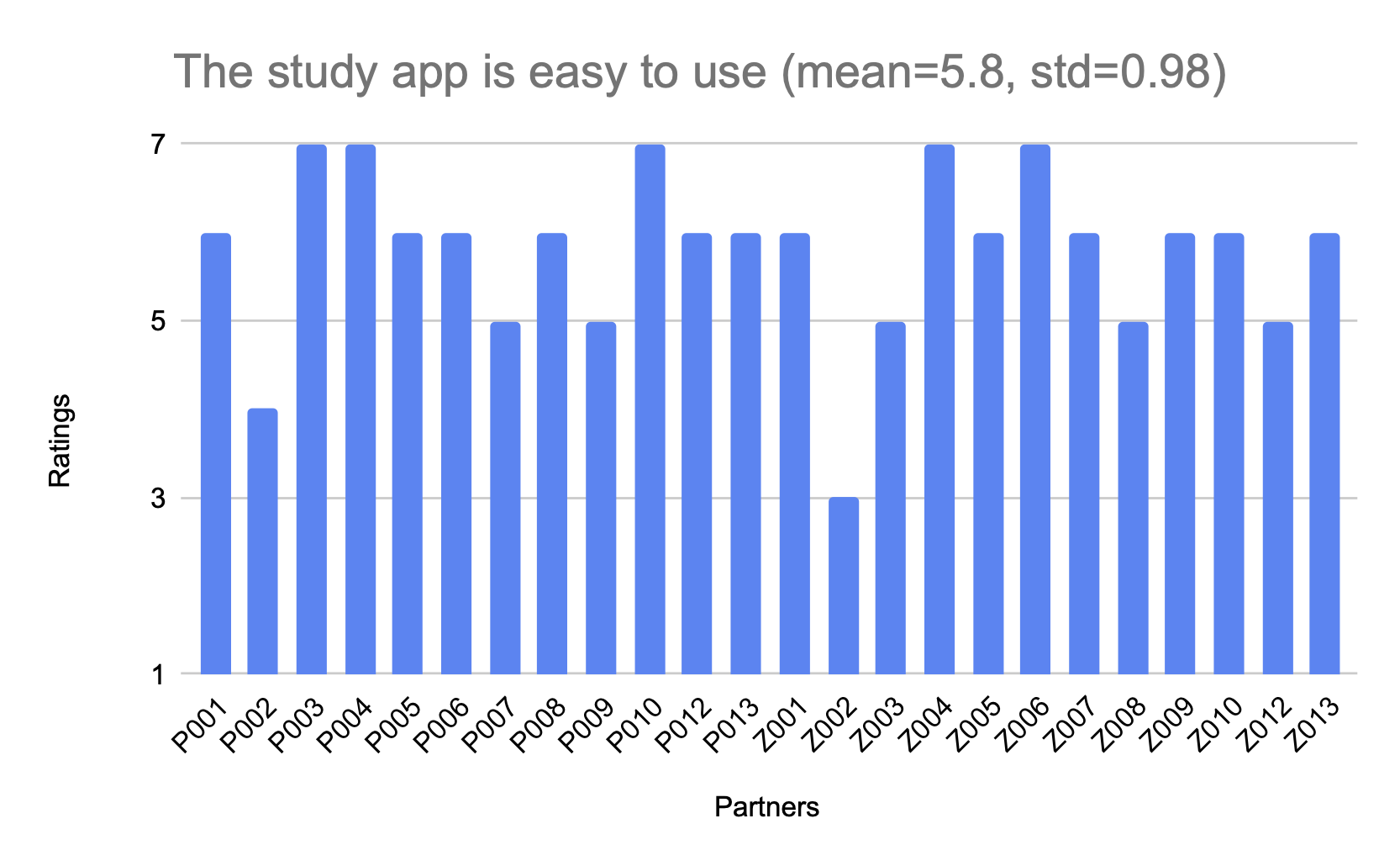}
\caption{Usability result showing partners’ responses to the statement “The study app was easy to use” on a 7-point Likert scale ranging from strongly disagree (1) to strongly agree (7) } 
\label{fig:usability_result}
\end{figure}

At the end of the 7-day field study, partners completed a self-report on their experience with the DyMand apps. They responded to the statement “The study app was easy to use” on a 7-point Likert scale ranging from strongly disagree (1) to strongly agree (7). As shown in Figure~\ref{fig:usability_result}, most ratings were high with a mean of 5.8 (std=0.98, N=24), which shows the DyMand system was easy to use. Also, partners wrote open-ended responses about potential areas of improvement for the DyMand system. One key recurring theme was that there was a delay in the self-report showing up as mentioned before. Another common suggestion was to reduce the number of questions.

\subsection{Errors}
On our server, when there were too many user intent messages (messages that indicate a recording is done in the watch), DeepStream could not maintain the connection between  the server and the client — a socket hang-up error. Though we could not find a technical solution to the problem, we managed to fix this by restarting the Docker container of DeepStream. As this happened only a couple of times during the study and as we found out the problem both times quickly during our periodic monitoring of the system, this error had a negligible effect on the data obtained. 

There were BLE connection issues, too. In the smartwatch, we found that after a few successful BLE connections, the smartwatches could not connect anymore with each other. We fixed this by periodically destroying and recreating the whole Bluetooth stack and not just the BLE services (scanning and advertising). 

\subsection{Challenges}
\label{sec:challenges}
Unavailability of Internet access sometimes resulted in self-reports not being shown on the smartphone in time as described before. This issue is a major limitation of the MobileCoach platform as the mobile app has to send a request to the server to then send a request back to the phone to then show the self-report on the screen introducing several points of failure. When this happens, the corresponding audio recording gets deleted as it is not valid for us to analyze without a corresponding self-report. The app then tries to record another sample and trigger another self-report, increasing the burden of self-report completion. A solution during the study would have been an offline trigger that syncs when the internet connection is restored. We did not anticipate this as we gave our participants phones with SIM cards for Internet access. Due to resource constraints, we could not change this later during the study. A future extension of the MobileCoach platform with an in-app offline trigger to show the self-report will circumvent these multiple pathways and will prevent this issue.

MobileCoach is designed such that the DeepStream server (section \ref{sec:mc-backend}) syncs the messages designed in the intervention designer to the app only when the app is open and used by the user. After many “recording done” user-intent messages to the server, the server triggers many self-reports and when the app is opened after an inactive period, it takes some time for the messages to sync with the DeepStream server. This issue caused some delay for the most recent message to appear when the app was opened. As the solution would have required major customizations in the MobileCoach framework, we did not address this issue.

Making the smartwatch app run continuously (section \ref{sec:cont-running}) was quite challenging. In addition to the Foreground service and changing the battery optimization settings, one of the system apps in the Nokia 6.1 kept shutting down our app after a few hours. We spent a lot of time trying to figure out the root cause and fix it by disabling the respective system app (com.evenwell.powersaving.g3). This issue is important to note for future studies that plan to implement long-running background services on the Android platform.

\section{Limitations and Future Work}
\label{sec:limitations_future}
When issues happened during the study, we sometimes did not have enough information to infer the cause until the devices were returned. Even though we had various key logs that would have helped, we did not implement a way to transfer the logs from the watch to the phone, and then to our server for monitoring. The DyMand system was already quite complex and we had time constraints, and hence, we decided not to invest the time and effort into implementing this feature. One key implementation challenge was the fact that the watch did not have a direct internet connection and hence, any such data transfer would have had to happen through the phone and then to the server, further introducing potential points of failure. To enhance better error analysis in the field, in the future, the DyMand system should be extended to transfer the logs from the watch to the phone and then to an external server.

Our algorithm for capturing couples’ interaction moments though performed well, it did not address some edge cases as described in the evaluation section (section \ref{sec:conv_moments}). Future work will update the algorithm to check if the partners are wearing the smartwatch as part of the closeness and speech detection check by leveraging other sensor data such as acceleration. Furthermore, we will implement a speaker identification method to check if the speech is from either of the partners and avoid being triggered by other people or speech from the radio or TV. This implementation will entail training a speaker identification model to extract a speech embedding from each partner during the setup and use it for comparison in real-time on the smartwatch.

Our DyMand system currently only collects relevant sensor and self-report data but it does not perform real-time recognition of constructs that are relevant for understanding couples’ chronic disease management such as emotional well-being, social support, and CDC. Future work will augment the DyMand system with such capabilities.

The DyMand system is generic in that it is suitable not only for couples’ diabetes management but also for studies in the context of related diseases such as hypertension, or mental health disorders in which sensor and self-reports about emotional well-being and health behavior are relevant for disease management and health intervention designs. Given it is open source, it can be extended or adapted, and our novel method for capturing partners’ conversations/interactions can be used by other researchers to develop similar apps to collect data to understand various constructs among other dyadic constellations such as friendships, and sibling and parent-child dyads. Also, another potential research use case is to better understand communication patterns in-situ and performance measures of teams in organizations.

\section{Conclusion}
\label{sec:conclusion}
In this work, we developed, deployed, and evaluated the DyMand smartwatch and smartphone system that captures couples’ dyadic interactions in daily life in the context of chronic disease management. It consists of a smartwatch app, a smartphone app, built on top of the MobileCoach platform that collects sensor and self-report data that are relevant for chronic disease management on couples’ interaction/conversation moments. We deployed DyMand in a 7-day field study and collected 85 hours of data from 13 heterosexual romantic couples from the German-speaking part of Switzerland. Key challenges affected the system’s performance and usability such as software errors, poor Internet connectivity, and long self-report questionnaires. Nonetheless, our evaluations showed that the system had good performance in triggering the collection of the expected number of sensor and self-report data, and capturing couples' conversation moments, and it was easy to use. The DyMand system would enable social, health, and clinical psychologists to understand the social dynamics of couples in everyday life and for developing and delivering behavioral interventions for couples who are managing chronic diseases. Our system could be customized and extended to be used in other contexts besides chronic disease management such as couples’ daily dynamics more broadly, workplace interactions, and other dyad constellations such as parent-child and sibling-sibling or roommate dyads.

\begin{acks}
Funding was provided by the Swiss National Science Foundation (CR12I1\_166348/1).
\end{acks}

\bibliographystyle{ACM-Reference-Format}
\bibliography{refs}

\end{document}